\documentclass{article}
\usepackage{spconf,amsmath,graphicx}
\usepackage{amsthm}
\usepackage{amsfonts,amssymb}
\usepackage{subfigure}
\usepackage{booktabs}
\usepackage{multirow}
\usepackage{framed}
\usepackage{graphicx} 
\usepackage{caption}
\usepackage[colorlinks,
linkcolor=black,
anchorcolor=black,
citecolor=blue]{hyperref}

\title{Diffusion-aided Extreme Video Compression with Lightweight Semantics Guidance}

\begin{document}
%
\name{Maojun Zhang$^{*\dagger}$\thanks{This work was supported by National Natural Science Foundation of China (Grant No. 624B2133, 62301487, and 62293481), the National Key R\&D Program of China under Grant No. 2024YFE0200804, and Fundamental Research Funds for the Central Universities under Grant No. 226-2024-00069. This work was supported by UKRI under the projects AI-R (EP/X030806/1) and INFORMED-AI (EP/Y028732/1), and by the SNS JU project 6G-GOALS under the EU Horizon program (Grant Agreement No. 101139232). For the purpose of open access, the authors have applied a Creative Commons Attribution (CCBY) license to any Author Accepted Manuscript version arising from this submission. }, Haotian Wu$^{\dagger}$, Richeng Jin$^{*}$, Deniz Gündüz$^{\dagger}$, Krystian Mikolajczyk$^{\dagger}$}
\address{$^*$ College of Information Science and Electronic Engineering, Zhejiang University, Hangzhou, China\\
$^\dagger$ Department of Electrical and Electronic Engineering, Imperial College London, UK\\
}
\maketitle
\begin{abstract}
	Modern video codecs and learning-based approaches struggle for semantic reconstruction at extremely low bit-rates due to reliance on low-level spatiotemporal redundancies. Generative models, especially diffusion models, offer a new paradigm for video compression by leveraging high-level semantic understanding and powerful visual synthesis. This paper propose a video compression framework that integrates generative priors to drastically reduce bit-rate while maintaining reconstruction fidelity. Specifically, our method compresses high-level semantic representations of the video, then uses a conditional diffusion model to reconstruct frames from these semantics. To further improve compression, we characterize motion information with global camera trajectories and foreground segmentation: background motion is compactly represented by camera pose parameters while foreground dynamics by sparse segmentation masks. This allows for significantly boosts compression efficiency, enabling descent video reconstruction at extremely low bit-rates. The code will be made publicly available.
\end{abstract} 
\vspace{-3mm}
\section{Introduction}\label{sec:intro}
\vspace{-3mm}
Explosive growth of video data, driven by increasing resolutions (4K/8K), higher frame rates (120/240 FPS), and emerging applications such as AR/VR, large-scale surveillance, and autonomous driving, is imposing unprecedented demands on bandwidth and memory, making highly efficient video compression techniques urgently necessary. Effective compression must leverage both spatial redundancies within frames and temporal correlations across frames to enable compact representation of visual content and motion information. Over the past decades, a series of standardized codecs have been developed, evolving from H.264/AVC \cite{wiegand2003overview} to H.265/HEVC \cite{sullivan2012overview} and most recently H.266/VVC \cite{bross2021overview}. While these standards achieve remarkable compression efficiency, their modular and hand-crafted designs often introduce visual artefacts and constrain global optimization. To overcome these limitations, deep learning–based approaches have emerged, initially improving isolated components \cite{xu2018reducing} (e.g., residual prediction) and later advancing to fully end-to-end frameworks that directly optimize the rate-distortion trade-off \cite{lu2019dvc,lin2020m,mentzer2022vct}. With continuous progress in architectures and training strategies, recent models such as DCVC-FM \cite{li2024neural} are beginning to surpass traditional codecs, highlighting the potential of learning-based video compression.

\begin{figure}[t] 
	\centering
	\includegraphics[width=1\linewidth]{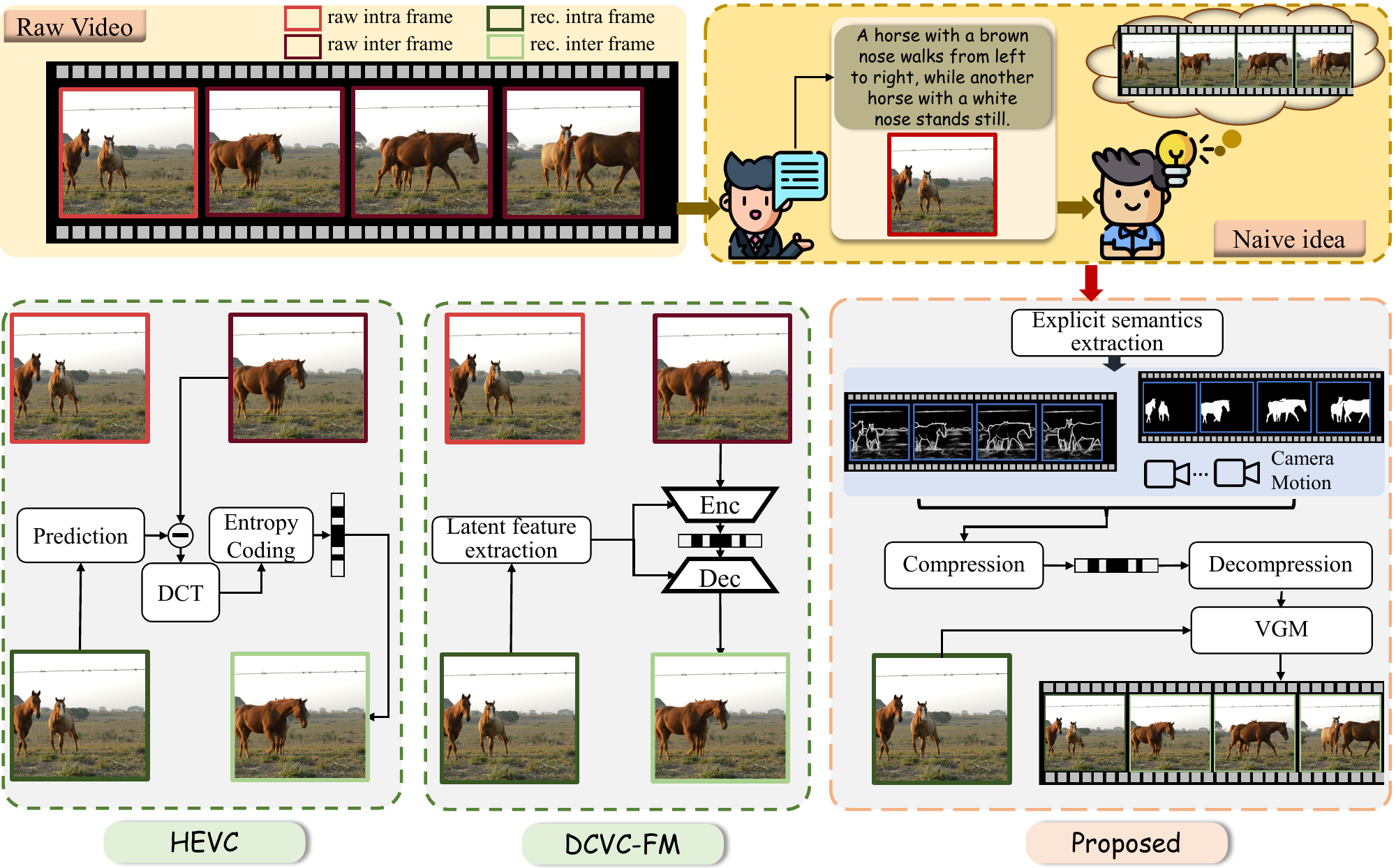}
	\vspace{-5mm}
	\caption{Comparison of the proposed framework with modern video codecs and learning-based compression framework.}\label{fig: paradigm comparison of compression methods}
\vspace{-7mm}	
\end{figure}
\begin{figure*}[t] 
	\centering
	\includegraphics[width=0.8\linewidth]{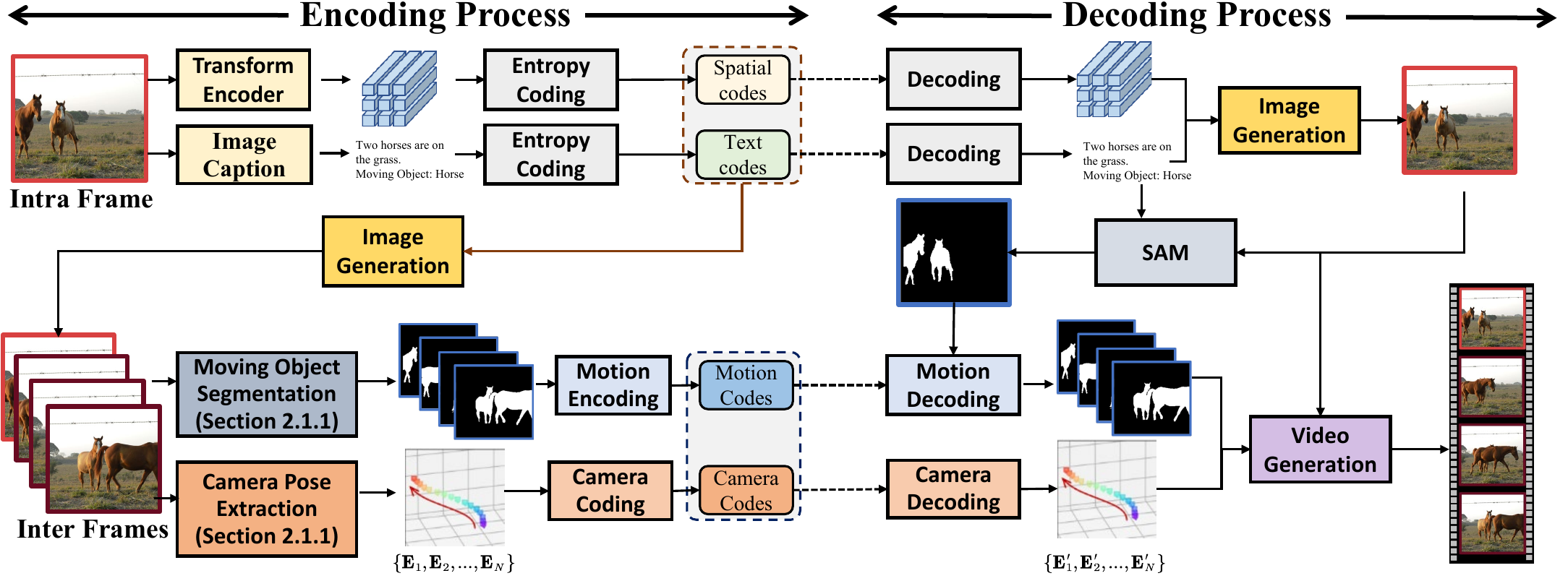}
	\vspace{-4mm}
	\caption{Extreme video compression framework via hierarchical motion semantics representation and compression}\label{fig: hierarchical semantics compression}
	\vspace{-7mm}
\end{figure*}
While modern codecs (e.g., H.265) and learning-based frameworks (e.g., DCVC-FM) have advanced video compression, the distortion–rate Pareto frontier remains far from optimal in extreme compression scenarios, especially when evaluated with semantic-level reconstruction metrics. This limitation is exemplified in Fig. \ref{fig: paradigm comparison of compression methods}, which illustrates this gap using a sequence with two horses, one moving and the other stationary. Existing codecs such as HEVC and DCVC-FM rely on pixel-level motion estimation \cite{sullivan2012overview,li2024neural}, which, though broadly effective, becomes inefficient under extreme compression as it fails to exploit coherent object-level motion shared across pixels of the same semantic entity. Notably, with only the first frame and a short text description, a human viewer could readily imagine the entire scene—implying that textual semantics may capture motion more compactly than raw pixel-level data. Realizing such an approach requires two components: (1) explicit semantic descriptors (e.g., text) that encode object-level content, and (2) a generative model capable of translating these descriptors into frames consistent with the specified semantics \cite{openai2024sora,blattmann2023stable,zhang2023i2vgen}. Similar concepts have been explored in image compression \cite{lei2023text+,careil2023towards,xu2025picd} and transmission 
\cite{zhang2025semantics}.  Extending this paradigm to video, however, is more challenging, as semantics must jointly represent both spatial structure and temporal dynamics. In particular, motion cues (e.g., optical flow) are difficult to express compactly as high-level descriptors, limiting direct extensions of image-based schemes. In this work, we propose a semantics-driven video compression framework that explicitly extracts and compresses motion semantics to enable extreme compression.
Our contributions are as follows.
(1) We introduce a motion representation framework that differentiates between background and foreground semantics. Background motion is compactly parameterized via estimated camera-pose trajectories \cite{he2024cameractrl}, while foreground motion is captured at finer granularity through temporally consistent segmentation maps. (2) We propose an in-context prompting pipeline that leverages video captions to identify moving objects, which are then used as prompts to the SAM2 model \cite{ravi2024sam} to generate accurate foreground masks. Camera-pose trajectories are extracted through standard pose estimation methods. (3) We fine-tune the Stable Video Diffusion-XL (SVD-XL) \cite{blattmann2023stable} model to support dual control from camera-pose trajectories and segmentation maps, enabling video reconstruction that preserves semantic consistency across space and time.

\vspace{-5mm}
\section{Methodology}
\vspace{-2mm}
Fig. \ref{fig: hierarchical semantics compression} provides a high-level overview of our proposed method. Given a raw video sequence, our objective is to reconstruct a sequence that remains semantically consistent with the original while incurring extremely low coding cost. Following the standard compression paradigm, we process the video fragment-by-fragment, where each fragment is defined as a group of pictures (GoP) $\mathbf{X}=\{\mathbf{x}_1,\mathbf{x}_2,\dots,\mathbf{x}_N\}$. The first frame of each GoP, referred to as the intra frame, is independently encoded using a conventional image compression technique. The remaining inter frames are encoded by leveraging motion information relative to the intra frame. 
Our focus in this paper is the compression of inter frames. 
We adopt a diffusion-based compression codec in \cite{li2024towards} to compress the intra frame. 
Unlike traditional approaches that directly infer motion with neural networks, our framework first extracts explicit semantic representations from the raw sequence. These semantic descriptors are then encoded, decoded, and finally employed to guide the image-to-video generation process. We shall detail each part as follows. 
\vspace{-3mm}
\subsection{Encoding}
\vspace{-2mm}
\subsubsection{Semantics Extraction}
\vspace{-2mm}
For a camera video, 
the foreground refers to regions where the camera is focusing on, while the background corresponds to static elements in the scene that are less dynamic. We represent these regions with different levels of granularity: a coarse representation for the background and a finer one for the foreground, reflecting their semantic significance.

\textbf{Background Motion Characterization:}
Background presents the overall scene, and its motion is generally simple: it reflects a global spatial transformation of the scene rather than rich, object-internal dynamics. Under extreme compression constraints we therefore do not represent background motion at the pixel level; instead, we model it as the overall scene movement induced by camera motion that can be characterized by the change of camera pose.   
Specifically, the camera pose is defined by its intrinsic and extrinsic parameters, denoted by $\mathbf{K}\in \mathbb{R}^{3\times 3}$ and $\mathbf{E}=[\mathbf{R};\mathbf{t}]$, respectively. The extrinsic parameters $\mathbf{E}$ define the position $\mathbf{t}$ and orientation $\mathbf{R}$ of the camera, while the intrinsic parameters $\mathbf{K}$ define the camera image characteristics. We use the FlowMap method in \cite{smith2024flowmap} to extract the camera pose, which computes high-quality camera poses for each frame through gradient descent optimization.  Each frame has independent extrinsic parameters ($\mathbf{E}_k$, $k$ is the frame index), while all the frames share a common $\mathbf{K}$, where only $4$ elements (focal length, image size) need to be determined. 
Therefore, to encode the motion of the background, only $12N+4$ floating-point numbers are needed, which is much more compression-efficient than the optical flow of each frame. 

\textbf{Foreground Motion Characterization:}
In addition to camera motion, foreground objects often exhibit complex, nonrigid motions (e.g., a person dancing or a horse running) that cannot be fully captured by a single global scene transform. To compactly represent such motion under extreme bit-rate constraints, we extract a per-frame foreground segmentation mask and use the motion of that mask-video to represent foreground motion. We choose segmentation maps because they deliver the most salient semantic motion information (what moves and how) while remaining highly compressible compared with richer geometric or contour-based modalities. Moreover, recent video-aware segmentation systems such as Segment Anything Model 2 (SAM2)  \cite{ravi2024sam} can produce consistent object masks for many practical scenarios, allowing the compressor to concentrate bits on the most relevant moving regions rather than the mask inconsistencies.

\begin{figure}[t] 
	\centering
	\includegraphics[width=1\linewidth]{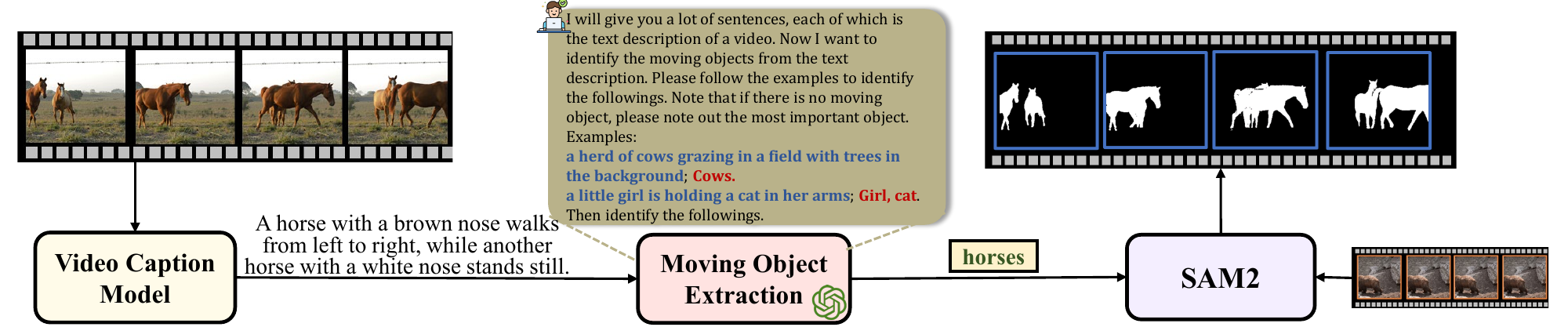}
	\vspace{-6mm}
	\caption{Foreground moving objects segmentation via in context learning with large language model.}\label{fig: foreground segmentation}
	\vspace{-6mm}
\end{figure}
Accurate identification of moving foreground regions, defined as areas containing the focused objects with independent motion, remains a major challenge especially for open-world video. Existing moving-object segmentation methods \cite{xie2024moving} often fail to generalize across unconstrained, diverse video content. To mitigate this, we propose an empirically effective segmentation pipeline (Fig. \ref{fig: foreground segmentation}) that combines semantic reasoning with instance-level segmentation. First, a captioning model generates a short textual description of the scene. A large language model (LLM) then parses this description to identify moving objects, leveraging its superior semantic comprehension and in-context learning abilities to resolve ambiguities that traditional BERT-based keyword extraction approaches fail to address. Finally, SAM2 is prompted with the LLM-derived object descriptions to produce precise, instance-level masks for the identified moving objects. This hybrid semantic–instance pipeline helps for focusing the available compression budget on the regions that matter most for perceived motion, improving both robustness and rate-efficiency in diverse, real-world videos. 
\vspace{-4mm}
\subsubsection{Semantics Compression}
\vspace{-2mm}

For the background motion, as mentioned, the camera pose trajectory is represented by the intrinsic and extrinsic parameters, with $4$ and $12N$ parameters, respectively. 
As the background motion is guided by relative camera motion, we work with relative poses. Specifically, we set the first frame as a unit camera pose, i.e., $\mathbf{R}_1=\mathbf{I}_{3\times 3}$, and $\mathbf{t}_1=\mathbf{0}_{3\times 1}$, and obtain the relative extrinsic of the remaining frames via $\mathbf{E}’_i=(\mathbf{E}_1)^{-1}\mathbf{E}_i$. Moreover, we observe that the camera pose trajectory $(\mathbf{E}’_1,…,\mathbf{E}'_N)$ is usually smooth and change slightly among frames. Given this, we first only keep poses for every other frame (i.e., $\mathbf{E}_2’,\mathbf{E}_4',...,\mathbf{E}_N'$) and then encode their differences, i.e., $\Delta\mathbf{E}_{1,2}', \Delta\mathbf{E}_{2,4}',...,\Delta\mathbf{E}_{N-2,N}'$, where  $\Delta\mathbf{E}_{i,j}=\mathbf{E}’_j-\mathbf{E}'_i$. Finally, we apply element-wise quantization: 
{
\setlength{\belowdisplayskip}{0pt}
\setlength{\belowdisplayshortskip}{0pt}
\vspace{-4mm}
\begin{align}
e_{i,j} = s_i*B_j+n_i,
\end{align}}
where $B_j\in\{0,...,255\}$ is the stored 8-bit integer, and $s_i$, $n_i$ are per-parameter scale and bias stored as 16-bit floats. 
The scale and bias are optimized over the set of extrinsic entries to minimize total quantization error. 
Then, we apply Huffman coding to the resulting bitstream. 

For the foreground segmentation map sequence, we employ DCVC-FM \cite{li2024neural} to compress the motion of foreground segmentation maps. 

\vspace{-5mm}
\subsection{Decoding}
\vspace{-3mm}
At the receiver side, we first reconstruct a half of camera pose $\mathbf{E}_2',\mathbf{E}_4',...,\mathbf{E}_N'$ and then reconstruct the remaining camera pose through interpolation. Specifically,  we utilize the spherical linear interpolation (Slerp) algorithm for reconstructing the rotation matrix, 
\vspace{-2mm}
{
\setlength{\belowdisplayskip}{-2pt}
\setlength{\belowdisplayshortskip}{-2pt}
\begin{align}
	\mathbf{R}_i'=q^{-1}\left(\frac{\sin((1-t)\Omega)}{\sin(\Omega)}q(\mathbf{R}_{i-1})+\frac{\sin(t\Omega)}{\sin(\Omega)}q(\mathbf{R}_{i+1})\right),
\end{align}
}
where $q(\cdot)$ and $q^{-1}(\cdot)$ denote the function for converting a rotation matrix to its corresponding quaternion representation and vice versa, respectively. $\Omega=\cos(q(\mathbf{R}_{i-1})^Tq(\mathbf{R}_{i+1}))$. 
For the translation vector, 
we adopt linear interpolation:
\vspace{-3mm}
{
\setlength{\belowdisplayskip}{0pt}
\setlength{\belowdisplayshortskip}{0pt}
\begin{align}
\mathbf{t}_{i}'=\frac{1}{2}\left(\mathbf{t}_{i-1}'+\mathbf{t}_{i+1}'\right).
\end{align}
}
\begin{figure}[t] 
	\centering
	\vspace{-2mm}
	\includegraphics[width=1\linewidth]{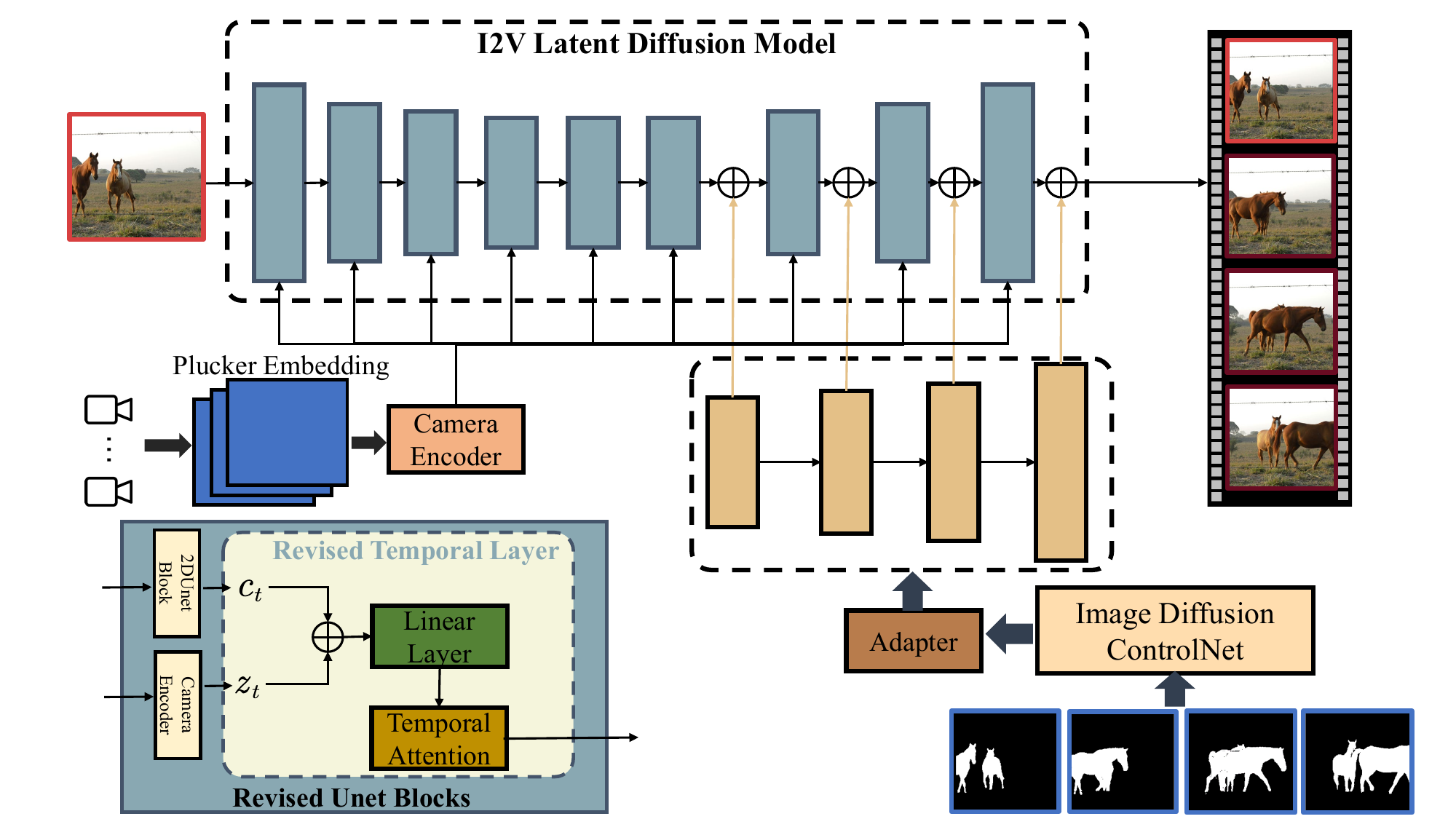}
	\vspace{-8mm}
	\caption{Extreme video compression framework via hierarchical motion semantics representation and compression.}\label{fig: camera pose and segmentation guided diffusion}
	\vspace{-6mm}
	\end{figure}
\begin{figure*}[!]
	\centering
	\setlength{\subfigtopskip}{0pt} 
\setlength{\subfigbottomskip}{-8pt} 
\setlength{\subfigcapskip}{-2mm} 
	\subfigure[LPIPS]{
		\label{Fig:awgn_wcsi:a} 
		\includegraphics[width=0.235\linewidth]{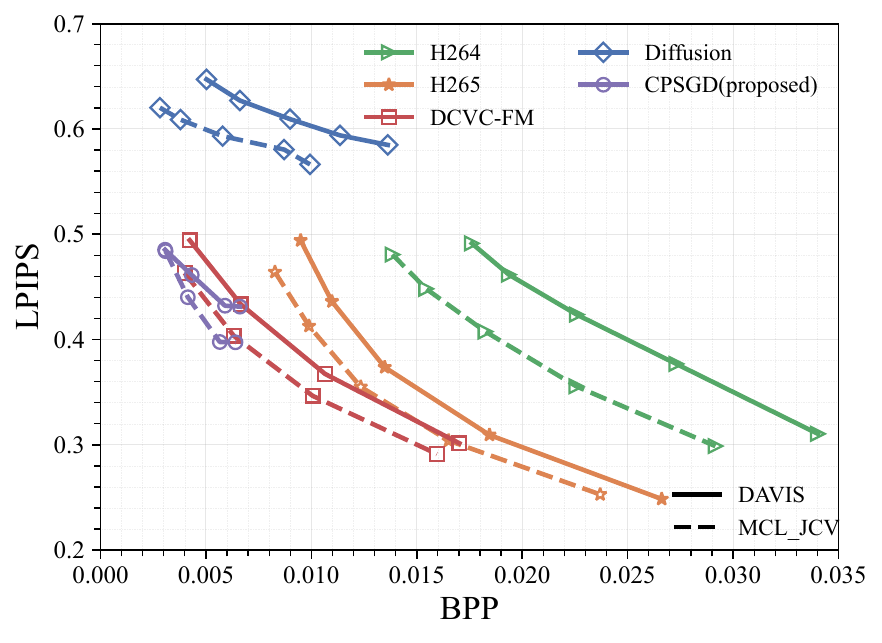}}
	\subfigure[FVD]{
		\label{Fig:awgn_wcsi:b} 
		\includegraphics[width=0.235\linewidth]{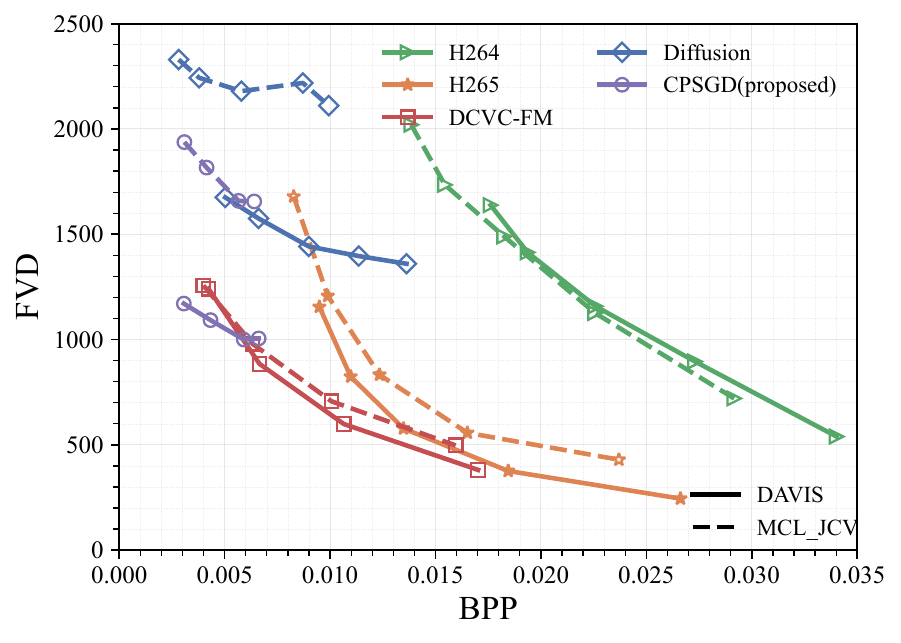}}
		\subfigure[CLIP Score]{
			\label{Fig:awgn_wcsi:c} 
			\includegraphics[width=0.235\linewidth]{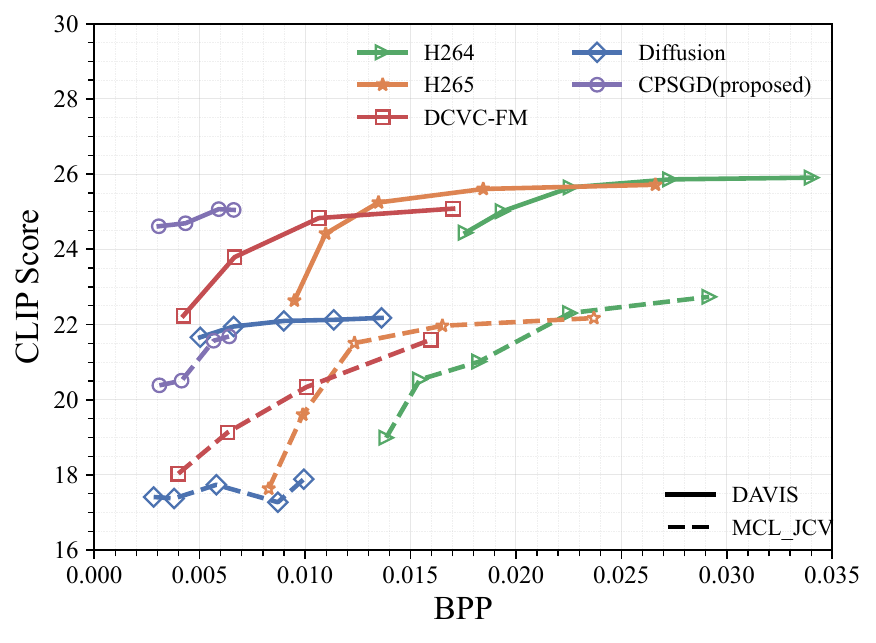}}
		\subfigure[Visual Quality \cite{huang2023vbench}]{
		\label{Fig:awgn_wcsi:d} 
		\includegraphics[width=0.235\linewidth]{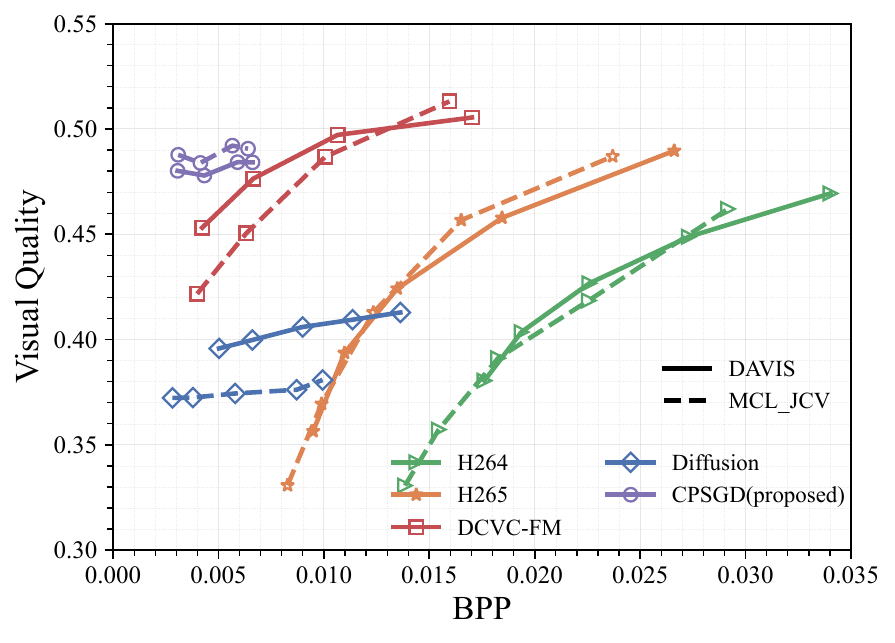}
		}
		\caption{Performance comparison of different compression schemes with different perceptual quality metrics.}\label{Fig:performance evaluation of different compression methods}
\vspace{-3mm}
	\end{figure*} 
\begin{figure*}[htbp]
\centering
\begin{minipage}[t]{0.68\textwidth}
\centering
\vspace{-22mm}
\captionof{table}{Compression analysis of different parts in the proposed CPSGD scheme.}\label{table: compression analysis of CPSGD}
	\resizebox{1\textwidth}{!}{
    \begin{tabular}{c|cr|cr|cr|cr|cr}
    \toprule[1.0pt] 
              & \multicolumn{4}{c|}{intra frame coding}    & \multicolumn{4}{c|}{inter frame coding}  & \multicolumn{2}{c}{\multirow{2}{*}{total  (BPP)}} \\ \cmidrule{2-9}
              & \multicolumn{2}{c|}{text (BPP)}    & \multicolumn{2}{c|}{spatial codes (BPP)} & \multicolumn{2}{c|}{camera pose (BPP)} & \multicolumn{2}{c|}{segmentation map motion} & \multicolumn{2}{c}{} \\ \midrule[1.0pt]
	Setting 1 & \multicolumn{1}{l|}{$1.02\times 10^{-4}$} &$3.33\%$  & \multicolumn{1}{l|}{$1.40\times 10^{-3}$}    & $45.69\%$   & \multicolumn{1}{l|}{$2.30\times 10^{-4}$}   &$7.50\%$    & \multicolumn{1}{l|}{$1.33\times 10^{-3}$}      &$43.48\%$      & \multicolumn{1}{l|}{$3.06\times 10^{-3}$} & 100\%     \\ \midrule
    Setting 2 & \multicolumn{1}{l|}{$1.02\times 10^{-4}$} & $2.36\%$ & \multicolumn{1}{l|}{$2.66\times 10^{-3}$}    & $61.57\%$   & \multicolumn{1}{l|}{$2.30\times 10^{-4}$}   & $5.31\%$   & \multicolumn{1}{l|}{$1.33\times 10^{-3}$}      &$30.77\%$     & \multicolumn{1}{l|}{$4.33\times 10^{-3}$}  & 100\%    \\ \midrule
    Setting 3 & \multicolumn{1}{l|}{$1.02\times 10^{-4}$} & $1.73\%$ & \multicolumn{1}{l|}{$4.24\times 10^{-3}$}    & $71.83\%$   & \multicolumn{1}{l|}{$2.30\times 10^{-4}$}   & $3.89\%$   & \multicolumn{1}{l|}{$1.33\times 10^{-3}$}      &$22.55\%$     & \multicolumn{1}{l|}{$5.91\times 10^{-3}$}  & 100\%    \\ \midrule
    Setting 4 & \multicolumn{1}{l|}{$1.02\times 10^{-4}$} & $1.54\%$ & \multicolumn{1}{l|}{$4.24\times 10^{-3}$}    & $64.15\%$   & \multicolumn{1}{l|}{$2.30\times 10^{-4}$}   & $3.47\%$   & \multicolumn{1}{l|}{$2.04\times 10^{-3}$}      &$30.83\%$     & \multicolumn{1}{l|}{$6.61\times 10^{-3}$}  & 100\%    \\
	\bottomrule[1.0pt]
    \end{tabular}}
\vspace{--5mm}
\end{minipage}
\begin{minipage}[!]{0.3\textwidth}
\centering
\setlength{\subfigtopskip}{0pt} 
\setlength{\subfigbottomskip}{0pt} 
\setlength{\subfigcapskip}{-2mm} 
	\subfigure[Origin]{
		\label{Fig:awgn_wcsi:a} 
{		\includegraphics[width=0.45\linewidth,trim=0 80 0 120,clip]{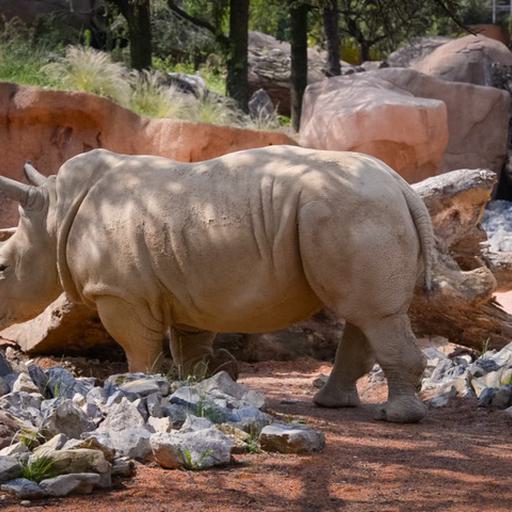}}
}	
		\subfigure[Diffusion(0.0077)]{
		\label{Fig:awgn_wcsi:d} 
		\includegraphics[width=0.45\linewidth,trim=0 80 0 120,clip]{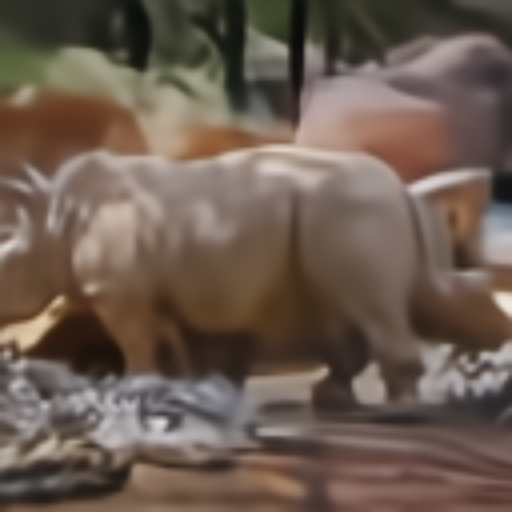}
		}
		\vspace{-4mm}
		\subfigure[{\fontsize{7pt}{8pt}\selectfont DCVC-FM}(0.0056)]{
		\label{Fig:awgn_wcsi:d} 
		\includegraphics[width=0.45\linewidth,trim=0 80 0 120,clip]{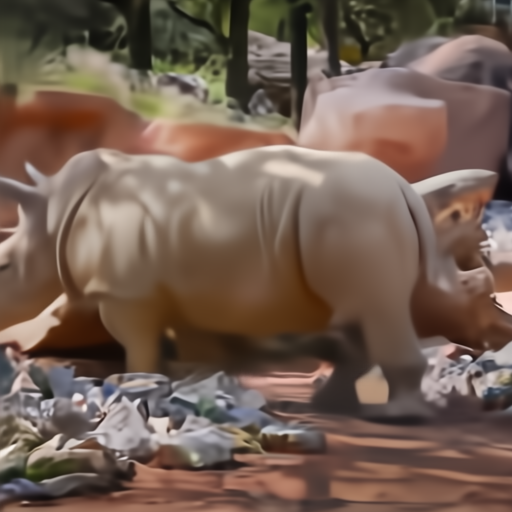}
		}
		\subfigure[CPSGD(0.0042)]{
		\label{Fig:awgn_wcsi:d} 
		\includegraphics[width=0.45\linewidth,trim=0 80 0 120,clip]{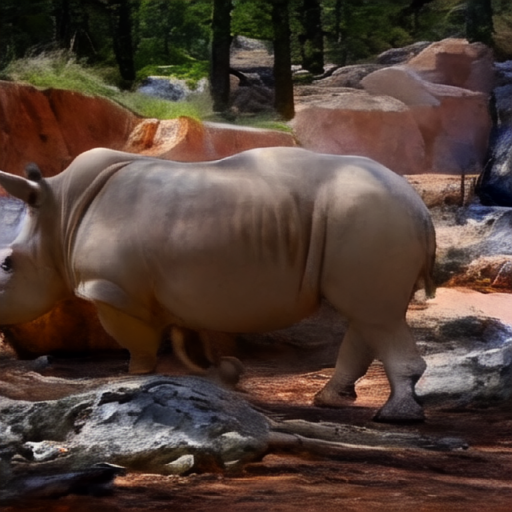}
		}
\vspace{1mm}
\captionof{figure}{Reconstruction example.}\label{fig: reconstruction example}
\end{minipage}
\vspace{-8mm}
\end{figure*}
In the video reconstruction stage, the receiver utilizes the decompressed foreground segmentation, camera pose trajectory, key frame, and text descriptions to reconstruct the video in the color space. To achieve this, a video diffusion model is required to generate the video based on the guidance from both the camera pose and foreground segmentation. However, such a diverse semantics guided diffusion has not been realized by existing methods. To fill this gap, we consider projecting both conditions into a pretrained I2V diffusion model \cite{blattmann2023stable} and training the corresponding adapter model. The architecture of the model is illustrated in Fig. \ref{fig: camera pose and segmentation guided diffusion}. 
For each frame, the camera pose is first transformed into its corresponding Plücker embedding, $\mathbf{P}_i \in \mathbb{R}^{6 \times H \times W}$, meaning that each pixel is represented by a vector $p_{u,v} = (\mathbf{o}\times\mathbf{d}_{u,v}, \mathbf{d}_{u,v})$, where $\mathbf{o} \in \mathbb{R}^3$ denotes the camera center in world coordinates, and $\mathbf{d}_{u,v} \in \mathbb{R}^3$ is a direction vector in the world coordinate system, pointing from the camera center to pixel $(u,v)$. This direction vector is given by 
\vspace{-3mm}
{
\setlength{\subfigtopskip}{-2pt} 
\setlength{\subfigbottomskip}{-2pt} 
\begin{align}
	\mathbf{d}_{u,v} = \mathbf{R} \mathbf{K}^{-1} [u,v,1]^T + \mathbf{t}.
\end{align}
}
The Plucker embedding sequence is then projected into the diffusion model. 
The noisy latent features and camera motion information are fused via an addition operation and a linear layer, after which the fused features are processed by temporal attention layers. 
On the other hand, as in \cite{lin2024ctrl}, the foreground motion information, derived from the segmentation map, is extracted using a pretrained image diffusion ControlNet model and then projected into the upsampling stage of the diffusion U-Net, where detailed spatial features are generated. 
\vspace{-9mm}
\section{Experiments}
\vspace{-4mm}
This section presents experiment results to evaluate the proposed camera pose and segmentation guided diffusion-based video compression (CPSGD). 
We use the SVD-XL as the base image-to-video model and introduce two adapters, one for camera-pose and one for segmentation guidance, which are each trained for 40,000 gradient steps and then concatenated and jointly fine-tuned on RealEstate10K. We compare CPSGD against H.264, H.265, DCVC-FM, and a diffusion-based compression baseline on DAVIS \cite{KhoRohrSch_ACCV2018} and MCL-JCV \cite{wang2016mcl} datasets; all videos are center-cropped and downsampled to $512\times512$. We use the bit-per-pixel (BPP) metric to characterize the compression rate. The reconstruction quality is measured with LPIPS, CLIP similarity, Frechet Video Distance (FVD), and 
the aesthetic quality in VBench \cite{huang2023vbench} to evaluate the visual quality. 

H.265 consistently outperforms H.264 and remains competitive with learning-based methods at moderate bit-rates (BPP$\geq0.01$). Below 0.01 BPP, however, 
the performance of H.265 drops sharply and is surpassed by learning-based schemes. DCVC-FM reaches a minimum of about 0.005 BPP and provides superior quality in this low-rate regime. Our CPSGD pushes the boundary further to  approximately 0.003 BPP, and at that bit-rate, yields better distortion and perceptual scores than a diffusion-based extreme compression baseline, producing visually more faithful reconstructions with fewer bits. These results demonstrate that generative, learning-based approaches offer substantial advantages under extreme compression constraints.

This subsection analyzes the bit-rate breakdown of the proposed CPSGD scheme on the DAVIS (Table \ref{table: compression analysis of CPSGD}). Across all four settings, text coding consumes under 4\% of the total bit-rate and mainly serves motion–semantic alignment. The segmentation maps and spatial codes require moderate bit allocation to ensure reconstruction fidelity. The four settings correspond to distinct operating points on the rate-distortion curve in Fig. \ref{Fig:performance evaluation of different compression methods}, where increasing bit budgets for encoding inter/intra frames enhances performance. Moreover, camera pose coding incurs negligible overhead (lower than 10\%), while foreground segmentation requires only $1.33\times10^{-3}$ BPP cost for three settings. This validates the superior compression efficiency of foreground and background-aware motion representation. 
A reconstruction example is provided in Fig. \ref{fig: reconstruction example}. 
\vspace{-6mm}
\section{Conclusions}
\vspace{-4.5mm}
In this work, we proposed a semantic-aware video compression scheme that explicitly models motion at different granularities: background motion is parameterized by camera pose, while foreground motion is represented by object segmentation maps. This separation enables compact, semantically meaningful coding that improves reconstruction quality under extreme bit-rate constraints. Empirical results demonstrated superior rate–distortion performance in the very low-BPP regime. Future work will explore even more efficient motion representations and joint end-to-end optimization of the generative compression pipeline.



\bibliographystyle{IEEEbib}
\bibliography{refs}

\end{document}